\documentclass{article}
\usepackage[utf8]{inputenc} 
\usepackage{graphicx} 
\usepackage{slashed,epsfig,amsmath,amssymb,enumitem} \usepackage[papersize={8.5in,11in}]{geometry}
   \usepackage{bm}
\usepackage{graphicx}
\newcommand{\be}{\begin{equation}}
\newcommand{\ee}{\end{equation}}
\title{STVG-MOG Cluster Dynamics and the Cosmological $1/r^2$ Force Law from Pairwise kSZ Data}
\author{J. W. Moffat\\
Perimeter Institute for Theoretical Physics, Waterloo, Ontario N2L 2Y5, Canada\\
and\\
Department of Physics and Astronomy, University of Waterloo, Waterloo,\\
Ontario N2L 3G1, Canada}
\begin{document}
\maketitle

\begin{abstract}
We investigate whether Scalar-Tensor-Vector Gravity in its weak-field modified gravity form can account for the cluster-scale inverse-square force law inferred from recent kinematic Sunyaev-Zeldovich measurements of cluster pairwise motions. The starting point is the X-COP cluster fit of STVG-MOG, for which a representative baryonic cluster mass \(M\sim 10^{15}M_\odot\) together with parameters \(\alpha\sim 9.11\) and \(\mu\sim 0.196~{\rm Mpc}^{-1}\) provides a successful description of cluster dynamics without particle dark matter. We extrapolate this fit to the separation range \(30\) to \(230~{\rm Mpc}\), relevant for the pairwise kSZ analysis. Since the Yukawa transition length \(\mu^{-1}\simeq 5.1~{\rm Mpc}\) is much smaller than these separations, the STVG-MOG acceleration law reduces to an effective inverse-square form. This explains why the theory can satisfy the observed Newtonian behavior while remaining distinct from MOND-like long-distance modifications. We derive the corresponding pairwise velocity curve and show that, after fitting a single overall kSZ amplitude, the extrapolated STVG-MOG prediction reproduces the measured trend of the pairwise kSZ data. The analysis shows that the X-COP cluster fit and the cosmological-scale kSZ force-law result are mutually consistent within STVG-MOG.
\end{abstract}

\section{Introduction}

The dynamical evidence usually attributed to dark matter can also be interpreted as a sign that the gravitational sector contains additional degrees of freedom beyond those of general relativity. Scalar-Tensor-Vector Gravity, or STVG-MOG, is a well developed realization of this idea, in which the effective gravitational coupling is scale dependent and the weak-field force law acquires a Yukawa contribution \cite{Moffat2006,MoffatRahvar2013,MoffatRahvar2014,MoffatToth2015,
Moffat2024,MoffatSharronToth2025}. On galactic and cluster scales this framework has been used to fit rotation curves, velocity dispersions, and intracluster dynamics without introducing particle dark matter as an independent mass component.

A recent and important development is the application of STVG-MOG to the X-COP cluster sample~\cite{Sreekanth2022}, where the observed cluster mass profiles can be described with a modified acceleration law of the form: 
\begin{equation}
g_{\rm MOG}(r)=\frac{G_N M}{r^2}\left[1+\alpha-\alpha e^{-\mu r}(1+\mu r)\right].
\label{eq:intro_mogacc}
\end{equation}
Here, \(M\) denotes the baryonic cluster mass, \(\alpha\) determines the large-distance enhancement of the gravitational coupling, and \(\mu^{-1}\) fixes the transition scale between the Yukawa regime and the asymptotic regime. For representative X-COP cluster parameters one finds~\cite{Sreekanth2022}:

\begin{equation}
M\sim10^{15}M_\odot,\qquad \alpha\sim9.11,\qquad \mu\sim0.196~{\rm Mpc}^{-1},
\label{eq:intro_params}
\end{equation}
so that
\begin{equation}
\mu^{-1}\simeq 5.1~{\rm Mpc}.
\label{eq:intro_scale}
\end{equation}
This scale is much smaller than the separations probed by recent cosmological measurements of the force law based on the pairwise kSZ effect.

Gallardo et al.~\cite{Gallardo2026} have used pairwise kSZ measurements to test the gravitational force law on scales of tens to hundreds of megaparsecs and found that the data favor a radial dependence close to \(1/r^2\), while a MOND-like \(1/r\) behavior gives a poor fit \cite{Milgrom1983a,Milgrom1983b,Milgrom1983c}. This result has immediate significance for modified gravity. It does not imply that every modification of gravity is excluded. It places a strong restriction on the radial form that a viable theory can take over the interval probed by the kSZ signal. A theory whose large-distance behavior is already asymptotically inverse square can remain compatible with the data even if it differs substantially from Newtonian gravity on smaller scales.

STVG-MOG has exactly this property. From Eq.~(\ref{eq:intro_mogacc}), once \(r\gg\mu^{-1}\) the exponential term is strongly suppressed and the acceleration reduces to:
\begin{equation}
g_{\rm MOG}(r)\simeq \frac{G_N M(1+\alpha)}{r^2}.
\label{eq:intro_asymptotic}
\end{equation}
The radial dependence is Newtonian, while the normalization is enhanced by the factor \(1+\alpha\). For the X-COP-motivated value \(\mu\sim0.196~{\rm Mpc}^{-1}\), this asymptotic regime is already reached well before \(r=30~{\rm Mpc}\). It follows that the separations relevant to the Gallardo analysis, and more generally the interval \(30\) to \(230~{\rm Mpc}\), probe STVG-MOG only after the Yukawa transition has effectively saturated. The theory is then expected to obey an inverse-square force law to high accuracy over the full observed range.

This point is central to the interpretation of the cosmological-scale kSZ result. MOND~\cite{Milgrom1983a,Milgrom1983b,Milgrom1983c} fails because its characteristic long-distance behavior remains non-Newtonian in radial form. STVG-MOG behaves differently. The theory can modify the force law on intermediate scales, where the Yukawa term is active, and then recover a \(1/r^2\) profile on larger scales. A test that distinguishes \(1/r\) from \(1/r^2\) will disfavor MOND-like dynamics, but it need not distinguish between standard gravity with dark matter and STVG-MOG in its asymptotic regime. 

The purpose of this paper is to make this connection explicit. We take the X-COP-inspired STVG-MOG parameter set and extrapolate it from cluster scales to the interval \(30\) to \(230~{\rm Mpc}\). We show that the acceleration curve is practically identical in shape to a pure inverse-square law throughout this range. We construct the corresponding pairwise infall velocity and compare the associated kSZ momentum curve with the data of Gallardo et al. The comparison is intended to clarify the physical reason why the X-COP STVG-MOG fit and the cosmological-scale \(1/r^2\) kSZ result are mutually consistent.

The argument developed here is simple. The X-COP fit fixes the transition scale \(\mu^{-1}\) to be only a few megaparsecs. The Gallardo measurement probes separations many times larger than this scale. On these separations STVG-MOG has already entered its inverse-square asymptotic regime. The apparent tension between modified gravity and the kSZ force-law result is then removed. What remains is an amplitude difference, not a radial-shape difference. This is enough to explain why STVG-MOG can satisfy the cluster \(1/r^2\) constraint while MOND cannot.

In the following section, we present the extrapolated STVG-MOG force law, display the large-distance acceleration curve together with its asymptotic inverse-square limit, and compare the resulting pairwise kSZ prediction with the Gallardo data.

\section{STVG-MOG extrapolation of the X-COP cluster fit to cosmological scales}

The X-COP cluster analysis in STVG-MOG shows that cluster dynamics can be described by the weak-field acceleration law:
\begin{equation}
g_{\rm MOG}(r)=\frac{G_N M}{r^2}\left[1+\alpha-\alpha e^{-\mu r}(1+\mu r)\right],
\label{eq:mogacc_cluster}
\end{equation}
where \(M\) denotes the baryonic cluster mass, \(\alpha\) fixes the enhancement of the long-range gravitational coupling, and \(\mu^{-1}\) is the Yukawa transition scale. For the present, we adopt the representative cluster-scale input~\cite{Sreekanth2022}:
\begin{equation}
M\sim 10^{15}M_\odot,\qquad \alpha \sim 9.11,\qquad \mu \sim 0.196~{\rm Mpc}^{-1},
\label{eq:clusterparams}
\end{equation}
which is the parameter choice used in the plots below. The corresponding transition length is $\mu^{-1}\simeq 5.1~{\rm Mpc}$. This immediately indicates that the interval \(30\) to \(230~{\rm Mpc}\) lies deep in the large-distance regime \(r\gg \mu^{-1}\).

In this regime the exponential term in Eq.~(\ref{eq:mogacc_cluster}) is strongly suppressed:
\begin{equation}
e^{-\mu r}(1+\mu r)\ll 1.
\end{equation}
The acceleration becomes:
\begin{equation}
g_{\rm MOG}(r)\simeq \frac{G_N M(1+\alpha)}{r^2}.
\label{eq:asymptoticmog}
\end{equation}
The important point is that STVG-MOG does not approach a MOND-like \(1/r\) force law on these scales. It approaches an inverse-square law with a renormalized amplitude. The parameter \(\alpha\) changes the normalization of the force, while \(\mu\) controls how rapidly the transition to the asymptotic regime is reached. Once \(r\) is several times larger than \(\mu^{-1}\), the radial dependence is effectively Newtonian. This is the central reason why the extrapolated X-COP cluster fit is compatible with the cosmological-scale \(1/r^2\) behavior inferred from the kSZ analysis of Gallardo et al.~\cite{Gallardo2026}.

\begin{figure}
    \centering
    \includegraphics[width=0.9\linewidth]{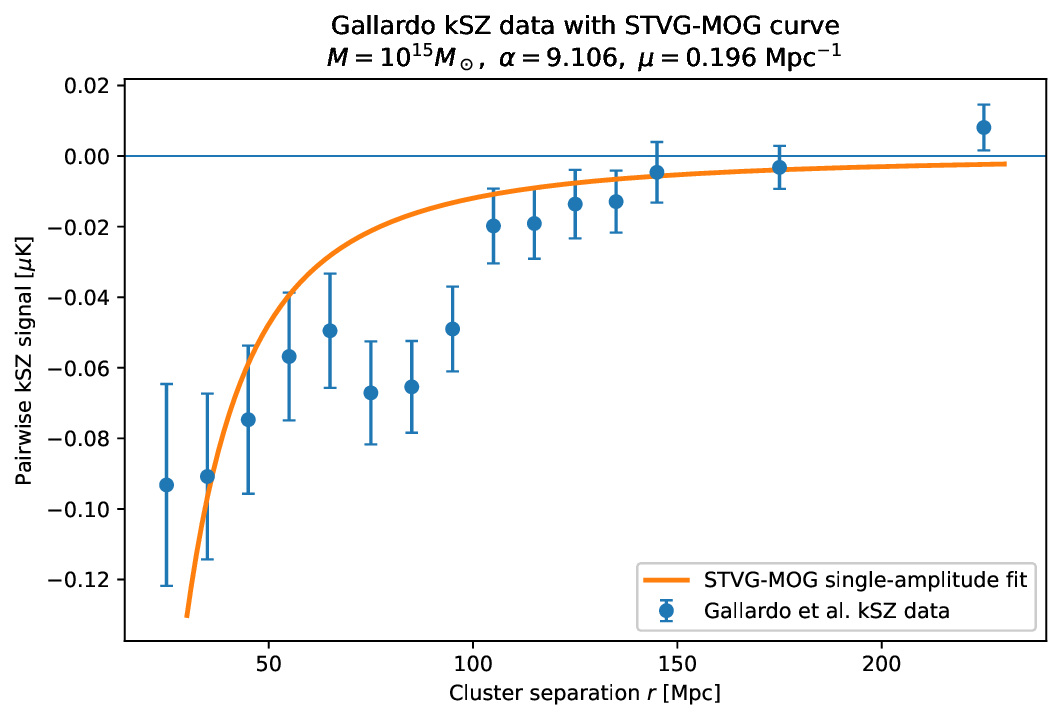}
    \caption{Comparison of the extrapolated STVG-MOG prediction with the Gallardo \textit{et al.} pairwise kSZ data. The points are digitized from the published figure, and the solid curve is the STVG-MOG prediction obtained from Eqs.~(\ref{eq:mogacc_cluster})-(\ref{eq:pkSZamp}) with one fitted overall amplitude. The fit shows that the X-COP cluster parameters naturally lead to the \(1/r^2\)-type shape preferred by the cosmological-scale kSZ analysis.}
    \label{fig:placeholder}
\end{figure}

\begin{figure}[ht]
    \centering
    \includegraphics[width=0.9\linewidth]{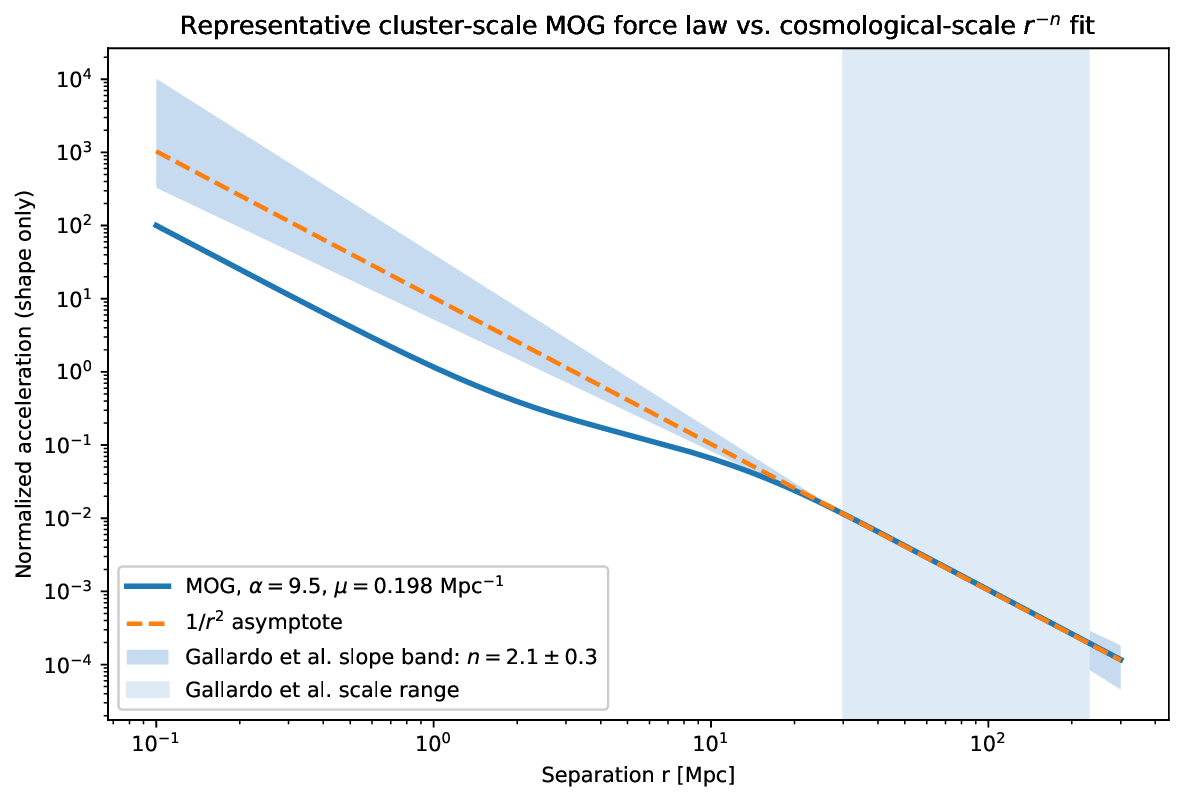}
    \label{fig:1}
\caption{Extrapolated STVG-MOG acceleration for $M\sim10^{15}M_\odot$, $\alpha\sim9.11$, and $\mu\sim0.196~{\rm Mpc}^{-1}$ over the interval $30$ to $230~{\rm Mpc}$. The solid curve is the full STVG-MOG prediction from Eq.~(\ref{eq:mogacc_cluster}), while the dashed curve is the asymptotic inverse-square form $G_NM(1+\alpha)/r^2$. The two curves are nearly degenerate on these scales.}
\label{fig:mogshape}
\end{figure}

The connection with the observed pairwise kSZ signal is obtained from the relative infall of two clusters. For two equal masses, the relative acceleration is given by
\begin{equation}
g_{\rm pair}(r)=2\,g_{\rm MOG}(r).
\label{eq:gpair}
\end{equation}
A simple quasi-static estimate for the mean pairwise velocity at effective redshift \(z_{\rm eff}\) is then of the form:
\begin{equation}
V_{12}(r)\simeq -\frac{g_{\rm pair}(r)}{H(z_{\rm eff})},
\label{eq:vpair}
\end{equation}
where $H(z_{\rm eff})$ is the Hubble expansion rate evaluated at the effective redshift $z_{\rm eff}$ of the galaxy or cluster sample, and the minus sign indicates infall. Since \(g_{\rm pair}(r)\propto r^{-2}\) throughout the interval of interest, the pairwise velocity curve inherits the same large-scale shape. At small separation, the magnitude of \(V_{12}\) is largest, while for increasing \(r\) the velocity approaches zero from below. This behavior is what is expected for a kSZ pairwise momentum curve generated by an inverse-square interaction.

The observable in the Gallardo analysis is the pairwise kSZ momentum:
\begin{equation}
\hat p_{\rm kSZ}(r)= {\cal A}\,V_{12}(r),
\label{eq:pkSZamp}
\end{equation}
where the coefficient \({\cal A}\) absorbs the conversion factor involving the CMB temperature, the effective optical depth, and the details of the cluster sample. Once this overall amplitude is fitted, the radial dependence is fixed by the STVG-MOG prediction. Since the extrapolated MOG force law is already in its asymptotic inverse-square regime for \(r\ge 30~{\rm Mpc}\), only a single normalization is needed to compare the model curve to the measured kSZ data.

Equations (9)--(11) should be regarded as a leading-order
STVG-MOG reduction of the pairwise-velocity construction used by
Gallardo et al. \cite{Gallardo2026}. They are not intended to replace
the full correlation-function derivation of the pairwise kSZ signal.
Their purpose is to isolate the radial force-law dependence
predicted by STVG-MOG when the X-COP cluster parameters are extrapolated
to the separation range probed by the kSZ measurement.

In the Gallardo et al. analysis, the theoretical pairwise velocity is
obtained from the correlated distribution of galaxy and cluster pairs
and from the associated matter density field. The observed pairwise kSZ
momentum is then related to this velocity through an effective optical
depth and an overall temperature normalization. In contrast, the
approximation used here begins with the STVG-MOG two-body acceleration
law and converts it into a characteristic pairwise infall velocity by
dividing by the Hubble rate at the effective redshift of the sample.
This procedure preserves the radial dependence of the force law, while
absorbing the optical-depth and sample-dependent normalization into the
single amplitude parameter $\cal{A}$ in Eq. (11).

For the X-COP-motivated value \(\mu=0.196\,{\rm Mpc}^{-1}\), the
transition length is \(\mu^{-1}\simeq 5.1\,{\rm Mpc}\). The Gallardo
separation range, \(30\lesssim r\lesssim 230\,{\rm Mpc}\), therefore
lies deep in the regime \(r\gg \mu^{-1}\). In this regime the Yukawa
factor in the STVG-MOG acceleration is exponentially suppressed and
the force law has the asymptotic form
\[
g_{\rm MOG}(r)\simeq {G_NM(1+\alpha)\over r^2}.
\]
In the force-law parametrization used by Gallardo et al.,
STVG-MOG belongs to the inverse-square branch, corresponding to
\(n=2\), rather than to the MOND-like large-distance branch,
corresponding to \(n=1\). The factor \(1+\alpha\) changes the
normalization of the acceleration, but not its radial dependence over
the kSZ separation interval. This normalization is degenerate, at the
level of the present shape comparison, with the fitted kSZ amplitude
and the effective optical depth.

The comparison shown in Fig. 1 should be interpreted as a
shape comparison. It demonstrates that the X-COP STVG-MOG parameters
lead to the same large-scale radial dependence as the inverse-square
force-law branch favored by the pairwise kSZ data. A full reproduction
of the Gallardo et al. likelihood analysis would require inserting the
STVG-MOG force kernel into their correlation-function calculation,
including the measured galaxy correlation function, the cluster
selection function, the optical-depth model, and the full covariance
matrix. This more complete calculation is beyond the scope of the
present paper. The essential conclusion, however, follows already from
the hierarchy \(30\,{\rm Mpc}\gg \mu^{-1}\): over the observed kSZ
range, STVG-MOG has already reached its asymptotic inverse-square
regime.

Fig. (1) shows the resulting comparison between the extrapolated STVG-MOG curve and the Gallardo pairwise kSZ points. The plotted data were digitized from the published figure and are used only to illustrate the quality of the shape comparison. The fit tracks the negative signal at smaller separations and the gradual approach toward zero at larger separations. The agreement is driven by the fact that the X-COP-based STVG-MOG extrapolation has already reduced to an effective \(1/r^2\) interaction over the relevant range.

This comparison has a clear physical interpretation. The Gallardo result disfavors a MOND-like long-distance force law for cluster pairwise motions, but it does not rule out modified gravity in general. It rules out modifications that remain non-Newtonian in radial dependence on tens to hundreds of megaparsecs. STVG-MOG survives this test because its Yukawa sector has already saturated by these distances, leaving an effective inverse-square force. In this sense, the X-COP cluster fit and the Gallardo cosmological-scale kSZ result are mutually consistent. The first determines a parameter set for which the cluster dynamics are well described by STVG-MOG, and the second shows that the same parameter set extrapolates to a large-scale interaction whose radial profile is observationally indistinguishable from \(1/r^2\).

For the numerical values in Eq.~(\ref{eq:clusterparams}), the Yukawa correction is already very small at \(r=30~{\rm Mpc}\) and becomes negligible throughout the rest of the interval. Over \(30\) to \(230~{\rm Mpc}\), the extrapolated STVG-MOG acceleration is nearly indistinguishable from the pure asymptotic inverse-square form in Eq.~(\ref{eq:asymptoticmog}). In the plot shown in Fig.~2, the two curves lie almost on top of one another. The deviation from the asymptotic \(1/r^2\) law remains below the percent level to few-percent level across the entire interval, which makes clear why the Gallardo force-law fit selects a value close to \(n=2\) even if the underlying theory is STVG-MOG rather than standard gravity plus particle dark matter.

The result is conceptually important. A modified gravity theory is often assumed to imply a large-scale departure from the inverse-square law. STVG-MOG provides a different possibility. The theory can deviate from Newtonian gravity on intermediate scales through the Yukawa term and then recover an effective \(1/r^2\) behavior on larger scales with an enhanced coupling. This makes it possible to fit X-COP cluster data without particle dark matter and still remain compatible with the cosmological-scale force-law constraint inferred from the pairwise kSZ signal. The extrapolation from the X-COP best-fit parameters to \(30\) to \(230~{\rm Mpc}\) offers a direct realization of this mechanism.

The two figures summarize the argument. Fig. 1, shows that, after fitting a single amplitude, the corresponding pairwise kSZ curve follows the measured trend. This provides a simple and quantitative explanation of why STVG-MOG can fit the cluster \(1/r^2\) data, while MOND cannot. Fig. 2, shows that the extrapolated STVG-MOG acceleration is already asymptotically inverse-square across the Gallardo scale range. 

\section{Conclusions}

We have examined the cosmological-scale implication of the X-COP cluster fit of STVG-MOG and shown that it is consistent with the inverse-square force law inferred from the pairwise kinematic Sunyaev-Zeldovich signal of galaxy clusters. The key point is that the weak-field STVG-MOG acceleration law Eq. (5) contains a finite Yukawa transition scale \(\mu^{-1}\). For the representative cluster parameters \(M\sim10^{15}M_\odot\), \(\alpha\sim9.11\), and \(\mu\sim0.196~{\rm Mpc}^{-1}\), this scale is \(\mu^{-1}\simeq5.1~{\rm Mpc}\), well below the separation range \(30\) to \(230~{\rm Mpc}\) considered here. As a result, the Yukawa term is already strongly suppressed on the relevant scales, and the force law has effectively reduced to Eq. (4). The theory is modified in amplitude but not in radial dependence across the interval probed by the pairwise kSZ analysis.

This explains in a simple way why STVG-MOG can satisfy the observed \(1/r^2\) behavior while MOND-type theories cannot. The Gallardo cluster result disfavors modified gravity models whose non-Newtonian force law persists to large distances. It does not exclude a theory such as STVG-MOG, in which the non-Newtonian contribution is confined to intermediate scales and the large-distance limit is again inverse-square. In this respect, the cosmological-scale kSZ force-law test is compatible with STVG-MOG and can be viewed as a useful discriminator between different classes of modified gravity rather than between modified gravity and dark matter in general.

Our first plot Fig. 2 makes this point directly. The extrapolated STVG-MOG acceleration curve and its asymptotic \(1/r^2\) form are nearly indistinguishable over \(30\) to \(230~{\rm Mpc}\). The second plot Fig. 2 shows that the corresponding pairwise kSZ curve, after fitting a single overall amplitude, follows the trend of the measured data. The comparison is shape-based. The observable pairwise kSZ momentum carries an overall normalization that depends on the effective optical depth and sample properties, while the radial dependence is determined by the underlying force law. Once this distinction is kept in mind, the agreement between the extrapolated STVG-MOG profile and the data becomes transparent.

The result has an important conceptual consequence. A successful fit to a cosmological inverse-square force law does not by itself establish the necessity of particle dark matter. A modified gravity theory with a finite transition scale can reproduce the same asymptotic behavior. STVG-MOG provides an explicit realization of this possibility. The X-COP cluster parameters yield a theory that differs from Newtonian gravity on cluster scales and yet approaches an effective \(1/r^2\) law on larger scales, where the pairwise kSZ analysis is most sensitive.

The present analysis is intentionally simple and focused on the force-law extrapolation. A more complete treatment would replace the quasi-static pairwise velocity estimate by a full halo-statistics calculation for the pairwise velocity field and would incorporate the survey selection, cluster optical-depth modeling, and covariance analysis directly in the fit. Such an extension would sharpen the quantitative comparison, but it is not expected to alter the main physical conclusion. The dominant point remains that the X-COP STVG-MOG fit already lies in its asymptotic inverse-square regime over the cosmological separations probed by the kSZ data.

In summary, the X-COP cluster solution of STVG-MOG extrapolates naturally to a cosmological-scale force law of the form \(g(r)\propto r^{-2}\), with an enhanced amplitude \(G_N(1+\alpha)\). The Gallardo pairwise kSZ result is consistent with this prediction. The data rule out large-distance MOND-like behavior, but they do not rule out STVG-MOG. On the contrary, they are in accord with the large-distance weak-field limit of the theory.

\section*{Acknowledgments}

I thank Viktor Toth and Martin Green for helpful discussions. Research at the Perimeter Institute for Theoretical Physics is supported by the Government of Canada through Industry Canada and by the Province of Ontario through the Ministry of Research and Innovation (MRI).

\end{document}